\documentclass[12pt]{article}

\begin{document}
\setcounter{page}{1}
\renewcommand{\thefootnote}{\fnsymbol{footnote}}
\pagestyle{plain} \vspace{1cm}
\begin{center}
\Large{\bf Black Hole Remnants in Hayward Solutions and
Noncommutative Effects}
\\
\small \vspace{1cm} {\bf S. H. Mehdipour$^{\rm
a,}$\footnote{mehdipour@liau.ac.ir}}\quad\quad and \quad\quad {\bf
M. H. Ahmadi$^{\rm a,}$\footnote{
ahmadi@liau.ac.ir}}  \\
\vspace{0.5cm} {\it $^{a}$Department of Physics, College of Basic
Sciences, Lahijan Branch, \\
Islamic Azad University, P. O. Box 1616, Lahijan,
Iran}\\

\end{center}
\vspace{1.5cm}
\begin{abstract}
In this paper, we explore the final stages of the black hole
evaporation for Hayward solutions. Our results show that the
behavior of Hawking's radiation changes considerably at the small
radii regime such that the black hole does not evaporate completely
and a stable remnant is left. We show that stability conditions hold
for the Hayward solutions found in the Einstein gravity coupled with
nonlinear electrodynamics. We analyse the effect that an inspired
model of the noncommutativity of spacetime can have on the
thermodynamics of Hayward spacetimes. This has been done by applying
the noncommutative effects to the non-rotating and rotating Hayward
black holes. In this setup, all point structures get replaced by
smeared distributions owing to this inspired approach. The
noncommutative effects result in a colder black hole in the small
radii regime as Hayward's free parameter $g$ increases. As well as
the effects of noncommutativity and the rotation factor, the
configuration of the remnant can be substantially affected by the
parameter $g$. However, in the rotating solution it is not so
sensitive to $g$ with respect to the non-rotating case. As a
consequence, Hayward's parameter, the noncommutativity and the
rotation may raise the minimum value of energy for the possible
formation of black holes in TeV-scale collisions. This observation
can be used as a potential explanation for the absence of black
holes in the current energy scales produced at particle colliders.
However, it is also found that if extra dimensions do exist, then
the possibility of the black hole production at energy scales
accessible at the LHC for large numbers of extra dimensions will be
larger. \\
{\bf Key Words}: Regular Black Holes, Hawking Temperature,
Noncommutative Geometry, Black Hole Remnant
\end{abstract}
\newpage

\section{\label{sec:1}Introduction}
Black holes (BHs) and singularities are accepted to be unavoidable
predictions of the theory of general relativity [1]. It is widely
believed that only a not yet attainable quantum gravity theory would
be capable to study the issue of central singularity of a BH
properly. However, various phenomenological approaches have been
considered in the literature in order to solve the problem of BH's
singularity with a regular center [2]. The Bardeen BH [3] is the
first regular model which has proposed as a spherically symmetric
compact object with an event horizon and without violating the weak
energy condition. The inside of its horizon is deSitter-like wherein
the matter has a high pressure. In 2006, the formation and
evaporation of a new kind of regular solutions was studied by
Hayward [4]. The static region of a Hayward spacetime is
Bardeen-like while the dynamic regions are Vaidya-like. A general
class of regular solutions utilizing a mass function that
generalizes the Bardeen and Hayward mass terms have been suggested
[5]. The authors of Ref.~[6] have discussed the massive scalar
quasinormal modes of the Hayward BH (H-BH). The motion of a particle
in background of a H-BH has been studied [7]. The accretion of fluid
flow around the modified H-BH has been investigated [8]. Recently,
the effects of thermal fluctuations on thermodynamics of a modified
H-BH have also been analyzed [9]. There have been a great number of
studies concerning regular BHs in the recent literature [10-14].

It was shown that the physical source of regular BHs can be
interpreted as the gravitational field of a nonlinear
electrodynamics (NED) [15-18]. The NED was founded by Born and
Infeld [19]. The NED theories emerge from low-energy effective
limits in specific models of string/M-theories [20-22]. There are
two basic aims in a NED theory. The first is to consider
electromagnetic field and particles within the context of a physical
source. The other great aim is to avoid letting physical quantities
become infinite. A similar procedure can be achieved by the NED
coupled to gravity in such a way that regular spherically symmetric
electrically charged solutions confirm the weak energy condition and
have an unavoidable deSitter centre. The regular BH solutions to
Einstein equations with physically reasonable sources have been
introduced by Ayon-Beato and Garcia [15-17]. In this model, the
Bardeen BH was reinterpreted as a magnetic solution to Einstein
equations with NED [18]. The regular BH solution in the $f(T)$
gravity coupled to NED has been found in [23]. The other solutions
of the combined Einstein and NED equations have also been reported
[24-26]. For considerably more details concerning the nonlinear
effects we suggest the following literature [27-34].

In addition, on the other hand, it is well-known that the appearance
of high energies in a noncommutative manifold is a consequence of
quantum fluctuation effects at very short distances wherein any
measurements to determine a particle position with an accuracy more
than an innate minimal length scale, namely the Planck length, are
hindered. Noncommutative BHs are naturally identified with the
possible running of this minimal length scale in BH physics. Based
on an inspired noncommutative model [35-39], instead of describing a
point particle as a Dirac-delta function distribution, it is
characterized by a Gaussian function distribution with a minimal
width $\sqrt{\theta}$, i.e. a smeared particle, where $\theta$ is
the smallest fundamental cell of an observable area in the
noncommutative spacetime, beyond which coordinate resolution is not
obvious. In this model, the energy-momentum tensor takes a new form,
while the Einstein tensor remains unchanged. As an important result,
the curvature singularity at the center of noncommutative BHs is
eliminated. This means that Planck scale physics may prevent the
appearance of a singularity in the center of a BH wherein a BH
remnant may be formed (for an extensive review of BH remnants, see
[40]).

In the group of various BH solutions, the rotating ones, without any
hesitations, are most suitable to fit the observational data proving
that collapsed objects display high angular momenta. The BH spin
plays a fundamental role in any astrophysical process. Hence, its
perfect comprehension is essential for the exact explanation of
astrophysical BHs, such as Cygnus X-1, utilizing their deviation
parameters from the Kerr BH [41-44]. Further, the astrophysical BHs
might be inherently quantum objects, macroscopically different from
the rotating ones predicted in Einstein's theory of gravity.
Furthermore, the rotating H-BH may be a good candidate for studying
how much quantum effects near the horizon can affect the radiation
released from the BH system. The gravitational detections at LIGO
[45-48], might be a clue for more efforts regarding the
gravitational wave from the models of regular BHs. Recently, the
upper bound thermally allowed in a head-on collision of two rotating
H-BHs was found by using the numerical method [49]. The author of
Ref.~[49] showed how much the gravitational radiation is dependent
on the parameters of the H-BH and found the effective range of the
parameters using the data from GW150914 and GW151226 [45-48]. In
this manner, the parameter $g$ of the H-BH is treated as a universal
constant in the spacetime, due to its relation to an energy level in
the near horizon of the BH. According to [49], assuming that the
mass of the first BH is unity and the second BH is smaller than the
first one, one finds that a large value of $g$ is not allowed and
the possible upper bound of the parameter $g$ ranges between $0.7$
and $0.8$. However, it needs more detection for the gravitational
wave generated by a BH binary because the analysis becomes more
precise in the limit of which a BH binary having a very small mass
ratio. As another important observable aspect is a study of the
gravitational lensing via regular BHs [50-52]. More recently, the
authors in [53] investigated the observables of a strong deflection
lensing, and estimated their values for the supermassive BH in the
center of our Galaxy (Sgr A$^\ast$). They have found that there is a
very high resolution beyond our current stage which is needed to
distinguish the modified H-BH from a Schwarzschild one.

It turns out to be a rather long process to solve Einstein's vacuum
equations directly for a rotating solution. Instead, by describing a
trick of Newman and Janis [54], one can obtain, for example, the
Kerr solution from the Schwarzschild case. The same trick can then
be applied to a regular case to achieve a rotating regular solution.
In 2013, Bambi and Modesto [55] apply the Newman-Janis algorithm to
the Hayward and to the Bardeen metrics to obtain a family of
rotating regular BHs. In this paper, we first consider the most
popular model of a regular BH derived in [4], namely the H-BH, and
then study its radiating behavior and the resulting remnant by
providing its Hawking temperature. We concisely study the dynamical
stability of static spherically symmetric exact solutions in a
self-gravitating NED theory via some conditions acting on the
electromagnetic Lagrangian which lead to the linear stability for
H-BHs solutions. In addition, the possibility of forming H-BHs at
energy scales of a few TeVs is studied by obtaining their remnant
mass. We compare different sizes of remnants with the noncommutative
ones by including the noncommutative corrections in the line element
of H-BH, i.e. the Noncommutative H-BH (NH-BH). Finally, using the
Newman-Janis algorithm which is often remarked as a short cut to
find spinning BH solutions via the corresponding non-rotating ones,
we consider again the inspired noncommutativity and determine the
Hawking temperature of the Noncommutative Rotating H-BH (NRH-BH).
Throughout the paper, natural units are used, i.e. $\hbar = c = G =
k_B = 1$ and Greek indices run from 0 to 3.

\section{\label{sec:2} Hayward solution}
The H-BH solution obtained by Hayward [4] is given by the following
metric,
\begin{equation}
\label{mat:1}ds^2=N(r)dt^2- N^{-1}(r)dr^2-r^2 d\Omega^2,
\end{equation}
with
\begin{equation}
\label{mat:2}N(r)=1-\frac{2m(r)}{r}=1-\frac{2Mr^2}{r^3+g^3}.
\end{equation}
where $g$ is a real free parameter and shows a positive constant
measuring the deviations from the standard Kerr spacetime. In the
above, the mass term $m(r)=\frac{Mr^3}{r^3+g^3}$ may show the mass
inside the sphere of radius $r$ such that in the limit $r\rightarrow
\infty$ it approaches the BH mass $M$. This solution is everywhere
nonsingular and the weak energy condition is not violated. The mass
term $m(r)$ interpolates between the de Sitter core and the
asymptotically flat infinity. The limits of large and small $r$ of
the metric function $N(r)$ are, respectively,
\begin{equation}
\label{mat:2.1}N(r)\approx1-\frac{2M}{r}+\frac{2Mg^3}{r^4},
\end{equation}
and
\begin{equation}
\label{mat:2.2}N(r)\approx1-\frac{2Mr^2}{g^3},
\end{equation}
which describes a central de Sitter solution, possibly in the regime
where quantum gravity effects should appear. Such a metric has event
horizons if $g^2\leq\frac{8}{9}2^{\frac{1}{3}}M^2$. So, for a given
value of $M$, quickly we find an upper bound for the Hayward
parameter, i.e., $g\leq g^*$. For example if we set $M=10$, there is
a critical value for $g$, namely $g^*\approx10.58$ that is the
condition for having one degenerate event horizon which means for
$g> g^*$ the horizons do not exist.

\begin{figure}[htp]
\begin{center}
\includegraphics{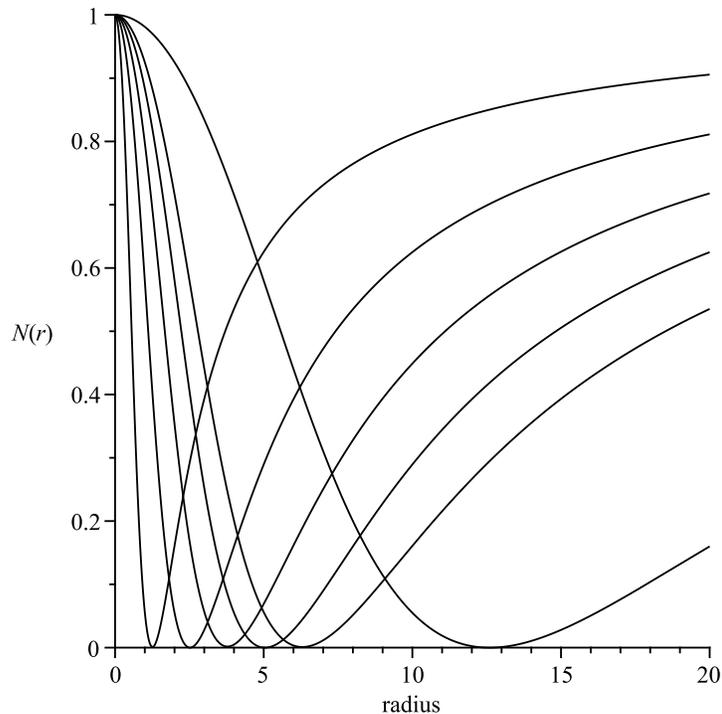}
\end{center}
\vspace{9 cm} \caption{\scriptsize {The temporal component of the
metric, $N(r)$, in terms of the radius $r$ for different values of
$g$. The figure displays the possibility of having extremal
configuration with one degenerate event horizon at a minimal nonzero
mass $M_0$. This presents the existence of $M_0$ such that the H-BH
may shrink to. On the right-hand side of the figure, from top to
bottom, the solid lines correspond to the H-BH for $g=1.00,~ 2.00,~
3.00,~ 4.00,~ 5.00,$ and $g=10.00$, respectively.}} \label{fig:1}
\end{figure}

The emitted feature of such a regular BH can now be simply analysed
by displaying the temporal component of the metric as a function of
radius for an extremal H-BH with different values of $g$. This has
been presented in Fig.~(\ref{fig:1}). This figure exhibits the
possibility of having an extremal configuration with one degenerate
event horizon at a minimal nonzero mass $M_0$. In fact, the
condition for having one degenerate event horizon is that $M=M_0$
which means for $M < M_0$ there is no event horizon. The existence
of a minimal nonzero mass may be interpreted as the deSitter-like
region corresponding to the interior of the horizon which yields a
remnant that the H-BH may shrink to.

\begin{figure}[htp]
\begin{center}
\includegraphics{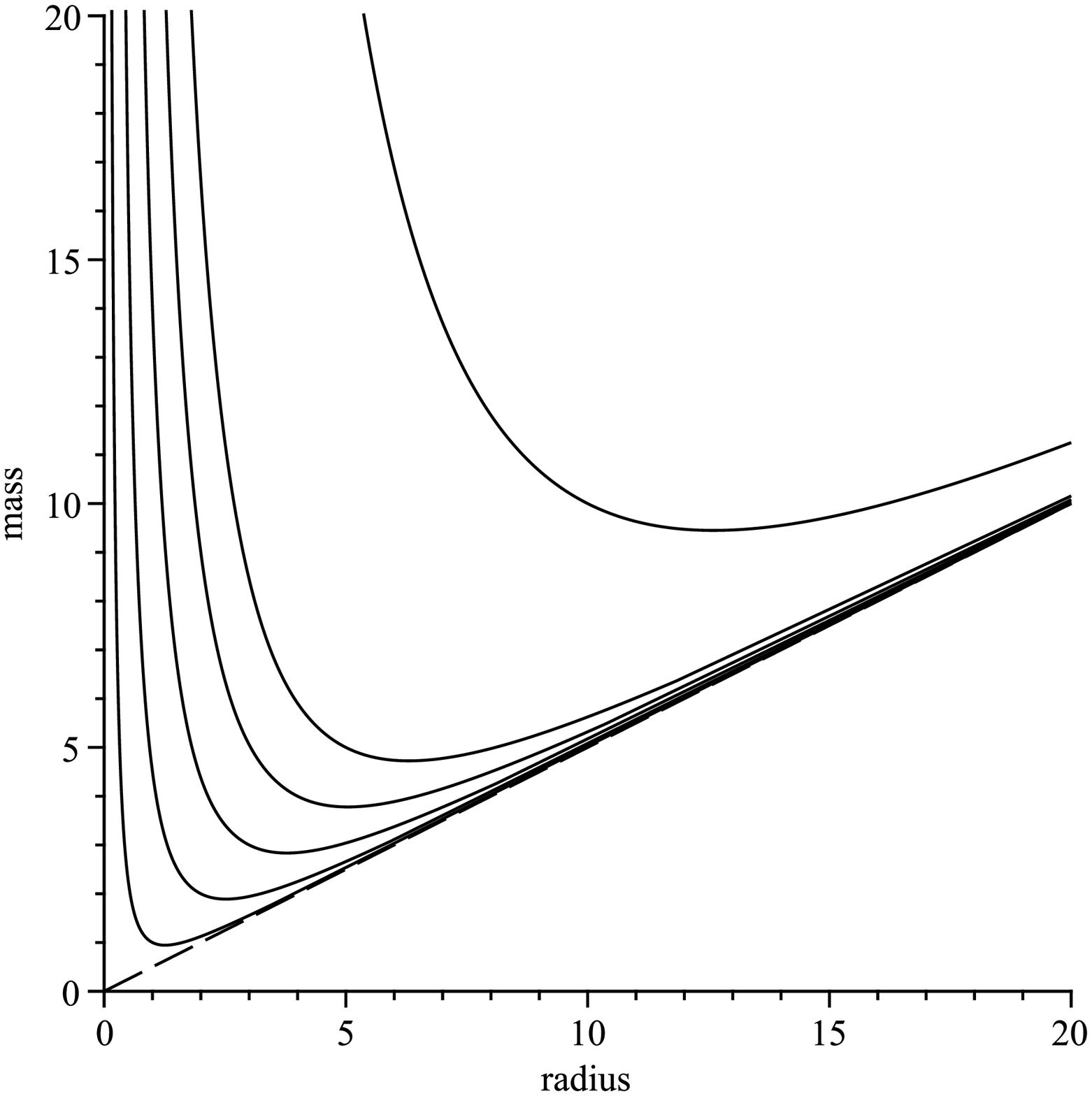}
\end{center}
\vspace{9 cm} \caption{\scriptsize {The mass of the H-BH as a
function of the horizon radius for different values of $g$. On the
left-hand side of the figure, from left to right, the solid lines
correspond to the H-BH for $g=1.00,~ 2.00,~ 3.00,~ 4.00,~ 5.00,$ and
$g=10.00$, respectively. The dashed line refers to the Schwarzschild
case so that it corresponds to $g=0$.  }} \label{fig:2}
\end{figure}

\begin{figure}[htp]
\begin{center}
\includegraphics{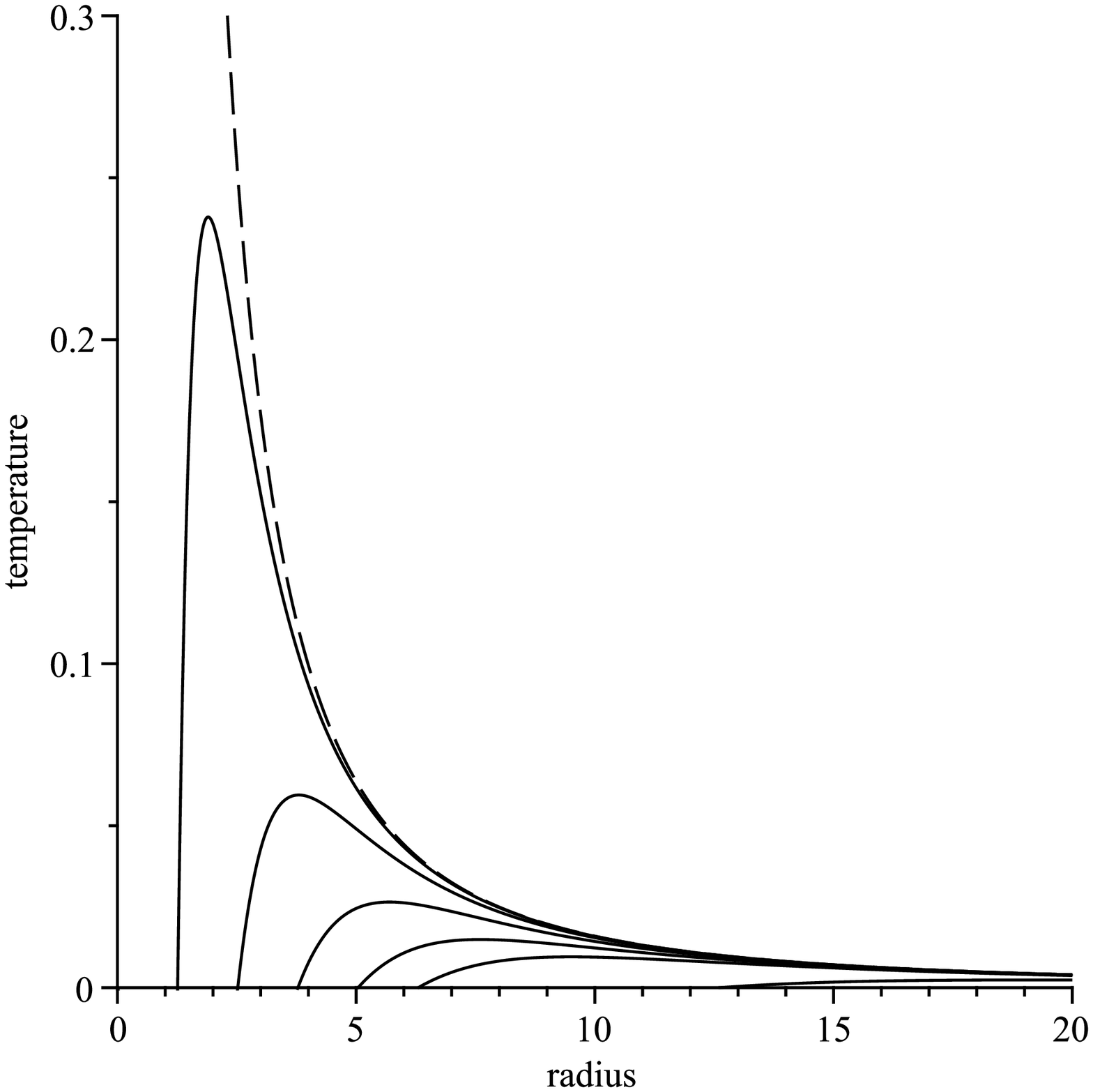}
\end{center}
\vspace{8.5 cm} \caption{\scriptsize {The Hawking temperature versus
the horizon radius. We have set $M=10.00$. On the left-hand side of
the figure, from left to right, the solid lines correspond to the
H-BH for $g=1.00,~ 2.00,~ 3.00,~ 4.00,~ 5.00,$ and $g=10.00$,
respectively. The dashed line refers to the Schwarzschild case so
that it corresponds to $g=0$.}} \label{fig:3}
\end{figure}

\begin{table}
\caption{\scriptsize {The remnant mass, the remnant radius and also
the maximum temperature of the H-BH for different values of $g$. As
the parameter $g$ increases the size and the mass of the H-BH
remnant increase but the maximum temperature decreases. For a large
amount of $g$, i.e. $g\gg1$, there is a linear relationship between
the remnant mass and the remnant radius. As can be seen from the
table, the results are confirmed by the numerical results of
Figs.~\ref{fig:1}, ~\ref{fig:2} and ~\ref{fig:3}.}}
\begin{center}
\begin{tabular}{|c|c|c|c|}
\hline
\multicolumn{4}{|c|}{H-BH} \\
\hline Free Parameter &   Remnant Mass & Remnant Radius  & Maximum Temperature \\
\hline$g=1.00$ & $M_0\approx0.94$ & $r_0\approx1.26$ & $T_H(max)\approx0.238$\\
\hline$g=2.00$ & $M_0\approx1.89$ & $r_0\approx2.52$ & $T_H(max)\approx0.059$\\
\hline$g=3.00$ & $M_0\approx2.83$ & $r_0\approx3.78$& $T_H(max)\approx0.026$\\
\hline$g=4.00$ & $M_0\approx3.78$ & $r_0\approx5.04$& $T_H(max)\approx0.015$\\
\hline$g=5.00$ & $M_0\approx4.72$ & $r_0\approx6.30$& $T_H(max)\approx0.009$\\
\hline$g=10.00$ & $M_0\approx9.45$ & $r_0\approx12.60$& $T_H(max)\approx0.002$\\
\hline
\end{tabular}
\end{center}
\label{tab:1}
\end{table}

The horizon radius of the H-BH can be obtained by the real positive
root of the following equation,
\begin{equation}
\label{mat:3}r_H^3-2Mr_H^2+g^3=0.
\end{equation}
So, one can find the H-BH mass in terms of $r_H$ as follows:
\begin{equation}
\label{mat:4}M=\frac{r_H^3+g^3}{2r_H^2}.
\end{equation}
The numerical results of the mass versus the radius are presented in
Fig.~\ref{fig:2}. As can be seen from Fig.~(\ref{fig:2}), the
minimal nonzero mass increases as the parameter $g$ increases. The
regularity at very short distances of the H-BH spacetime implies a
remnant mass corresponding to a remnant radius $r_0$. Here we have
shown that the final stage of the evaporation of H-BH is a remnant
in which it has an increasing size with raising its own free
parameter.

When such a regular BH radiates, its temperature is given by
\begin{equation}
\label{mat:6}T_H=\frac{1}{4\pi}\frac{dN(r)}{dr}\bigg|_{r=r_H}=\frac{Mr_H(r_H^3-2g^3)}{2\pi(r_H^3+g^3)^2}.
\end{equation}
Due to the emission of Hawking radiation, the Hawking temperature
finally reaches a peak at the final stage of the evaporation and
then abruptly drops to zero so that a stable remnant is appeared.
The remnant radius can be determined from $T_H=0$, namely
$r_0=2^{\frac{1}{3}}g$. This minimum radius corresponds to the
remnant mass $M_0=\frac{3}{2^{\frac{5}{3}}}g$.

The numerical result of the Hawking temperature in terms of the
horizon radius is displayed in Fig.~\ref{fig:3}. According to
Fig.~\ref{fig:3}, the temperature peak of the H-BH decreases as the
parameter $g$ increases, so a H-BH for a larger amount of $g$ is
colder and its remnant is bigger. If we set $g=0$, so the Hawking
temperature for the Schwarzschild BH, i.e. $T_H=\frac{M}{2\pi
r_H^2}$, which is accompanied by a divergence at $M=0$, is
retrieved.

Table~\ref{tab:1}, for further specifications of the H-BH remnant,
shows the numerical results of the remnant size, the remnant mass
and also the maximum temperature for different values of $g$. In
accordance with Table~\ref{tab:1}, as $g$ becomes larger both the
minimal mass and the minimal radius get larger but the temperature
peak becomes smaller. In the limit $g\gg1$, the free parameter $g$
is proportional to the remnant mass and to the remnant radius, i.e.
$g\propto M_0\propto r_0$. In other words, for an adequately large
amount of $g$ which corresponds to a large radius, there is a linear
relationship between the minimal mass and the minimal radius which
is similar to the result appeared in the relationship between the
horizon radius and the BH mass for the Schwarzschild BH.

Here we would like to check the thermodynamical stability of the
H-BH. The thermodynamic stability of a system can be investigated in
different ensembles. In the canonical ensemble, the thermal
stability of a system is determined by the sign of its heat
capacity. The positivity of the heat capacity is sufficient to
ensure thermal stability of a thermodynamical system. So, a BH is
thermodynamically unstable when its heat capacity is negative. The
heat capacity of the BH can be obtained using $C=\frac{\partial
M}{\partial r_H}\left(\frac{\partial T_H}{\partial r_H}\right)^{-1}
$. For $M>M_0$, we have $\frac{\partial M}{\partial r_H}>0$ (see
Fig.~\ref{fig:2}). Thus the sign of the heat capacity will be
determined by the sign of $\left(\frac{\partial T_H}{\partial
r_H}\right)^{-1} $ (see Fig.~\ref{fig:3}). There is a stable region
of positive heat capacity, which represents the near-horizon
thermodynamics. In this region, the temperature reaches a maximum
value of its amount at the position that the slope of the
temperature curve is zero, i.e. $\frac{\partial T_H}{\partial
r_H}=0$, then the heat capacity becomes singular for this special
value, known as Davies' point, where the temperature is maximum and
the heat capacity changes from negative infinity to positive
infinity [56, 57]. In this point the whole thermodynamic process
separates into two stages; the early stage with a positive heat
capacity and the late stage with a negative heat capacity. Indeed,
this is the process from an initial unstable large BH to a final
stable extremal BH.

It is obvious that an asymptotically flat uncharged BH is thermally
unstable, so in order to achieve a stable BH, one can add the
cosmological constant, the electric charge or the magnetic charge to
the solutions. In the next section, we briefly review a special
theory of NED, which predicts a recognizable physical source for the
central regularity of the H-BH, and indicate an exact solution for
the H-BH to analyze its stability in the NED theory.

\section{\label{sec:3}Stability analysis in an exact regular BH solution in Einstein-nonlinear
electrodynamics} It has been shown that general relativity coupled
to NED yields the nontrivial spherically symmetric solutions with a
globally regular metric [15-18]. The H-BH is also an exact solution
obtained in the Einstein gravity coupled with NED [25]. In this
section, we briefly study the H-BH in NED and show its stability
under linear perturbations. Stability properties in self-gravitating
NED were investigated by Moreno and Sarbach [58]. They found
adequate criteria for the linear stability with respect to arbitrary
linear fluctuations in the metric and in the gauge potential. These
criteria are in the form of inequalities to be fulfilled by the NED
Lagrangian density and its derivatives.

The action describing the dynamics of a self-gravitating NED field
in general relativity is
\begin{equation}
\label{mat:6.1}S=\frac{1}{4\pi}\int\left(\frac{R}{4}-L(F)\right)\sqrt{-\textsf{g}}d^4x,
\end{equation}
where $R$ is the Ricci scalar with respect to the spacetime metric
$\textsf{g}_{\alpha\beta}$, and the Lagrangian density $L(F)$
denotes a nonlinear function of the Lorentz invariant
$F=\frac{1}{4}F_{\alpha\beta}F^{\alpha\beta}$, where
$F_{\alpha\beta}=\partial_\alpha A_\beta-\partial_\beta A_\alpha$ is
the electromagnetic field. The Lagrangian density is an arbitrary
function which leads to $L(F)\approx F$ at small $F$, i.e. for the
weak field limit, describes the Maxwell theory. The temporal
component of Einstein equations, $G_0^{~0}$, resulting from the
above action yields
\begin{equation}
\label{mat:6.2}m(r)=\int L(F)r^2dr.
\end{equation}
Substituting $m(r)$ into a static and spherically symmetric
configuration one finally finds the Hayward metric [25].

For the stability analysis, it is convenient to consider the
Lagrangian density to be a function of the dimensionless variable
$y=\sqrt{2g^2F}=\frac{g^2}{r^2}$. Note that the parameter $g$ here
is not just a universal constant, but a magnetic charge associated
with a physically reasonable matter content. The H-BH which does
have a correct weak field limit is obtained from the Lagrangian
density
\begin{equation}
\label{mat:6.3}L(y)=\frac{3}{2sg^2}\frac{y^3}{(1+y^{\frac{3}{2}})^2},
\end{equation}
where $s=\frac{|g|}{2M}$ is a positive constant. The metric function
$N$ in terms of $y$ is given by
\begin{equation}
\label{mat:6.4}N=1-\frac{1}{s}\frac{y^{\frac{1}{2}}}{1+y^{\frac{3}{2}}}.
\end{equation}
The equation $N(y_m, s) = 0$ is solved by the single root $s=s_c =
\frac{2^{\frac{2}{3}}}{3}$, where $y_m=2^{-\frac{2}{3}}$ is a single
minimum of $N$. At $y_m$, for $s < s_c$ the minimum of $N$ is
negative, for $s = s_c$ the minimum vanishes and for $s > s_c$ the
minimum is positive. In other words, for $g^2<
4M^2s_c^2=\frac{8}{9}2^{\frac{1}{3}}M^2$ we have two event horizons,
for $g^2=\frac{8}{9}2^{\frac{1}{3}}M^2$ the horizons shrink into a
single one (extremal BH), and no event horizon for $g^2>
\frac{8}{9}2^{\frac{1}{3}}M^2$.

According to [58] (see also [59]), one can conclude that the linear
stability criteria on the corresponding BH solutions obligates the
satisfaction of the following inequalities
\begin{equation}\label{mat:6.5}
\begin{array}{ll}
L>0,\\
L_{,y}=\frac{9}{2sg^2}\frac{y^2}{(1+y^{\frac{3}{2}})^3}>0,\\
L_{,yy}=\frac{9}{4sg^2}\frac{4y-5y^{\frac{5}{2}}}{(1+y^{\frac{3}{2}})^4}>0,\\
3L_{,y}\geq yNL_{,yy},\\
\end{array}
\end{equation}
where $L_{,y}$ and $L_{,yy}$ are the first and second derivatives of
$L$ with respect to $y$, respectively. The last inequality above can
also be written as
\begin{equation}
\label{mat:6.6}3\geq Nf(y)>0,
\end{equation}
where the function $f(y)$ is defined as
\begin{equation}
\label{mat:6.7}f(y)=\frac{yL_{,yy}}{L_{,y}}=\frac{4-5y^{\frac{3}{2}}}{2(1+y^{\frac{3}{2}})}.
\end{equation}
The function $f(y)$ is smoothly lessening with $f(0) = 2$. Also,
since the metric function has a single minimum at
$2^{-\frac{2}{3}}$, therefore one has $y_H\leq2^{-\frac{2}{3}}$,
where $y_H$ is the value of $y$ at the event horizon. Accordingly,
it is easy to check that the conditions (\ref{mat:6.5}) for all
$0\leq y\leq y_H$, are satisfied and so Hayward BHs are stable.

\section{\label{sec:4}Higher-dimensional Hayward solution}
In the continuing search for quantum gravity, the BH thermodynamics
may be associated with future experimental results at the LHC
[60-62]. For example, the semiclassical analysis of loop quantum BHs
prepares regular BHs without singularity such that their minimum
sizes are at the Planck scale regime [63]. As another example,
gravity's rainbow motivated by doubly special relativity, using the
modified dispersion relation [64], produces remnants at the final
phase of the BH evaporation. As an important note, the
thermodynamics descriptions of H-BHs are substantially similar to
that of the framework of gravity's rainbow [65]. In the context of
gravity's rainbow [66], the remnant mass has found to be greater
than the energy scale at which experiments were performed at the
LHC. In this section, we shall extend our study into extra
dimensions to investigate the phenomenological implications on the
production of BHs at TeV scales.

The metric (\ref{mat:1}) can be generalized to a higher-dimensional
spacetime. Considering static, spherically symmetric $d$-dimensional
spacetime one obtains [67]
\begin{equation}
\label{mat:6.8}ds^2=Ndt^2- N^{-1}dr^2-r^2 d\Omega_{d-2}^2,
\end{equation}
with
\begin{equation}
\label{mat:6.9}N=N(r,r_g)=1-\frac{r_g^{d-3}r^2}{r^{d-1}+g_d^3},
\end{equation}
where $g_d=r_g^{\frac{d-3}{3}}l^{\frac{2}{3}}$ is Hayward's
parameter in the higher dimensional spacetime and depends on the
extra dimension models. The parameter $r_g$ is the gravitational
radius of the H-BH in extra dimensions in which at far distance it
reproduces its correct Schwarzschild asymptotic form in the
4-dimensional case, i.e. $2M$. The parameter $l$ is a length-scale
parameter. One of the main assumptions here is that there is a
critical energy $\Lambda$ and the corresponding length-scale
parameter $l$ in such a way that one has $l=\Lambda^{-1}$. This
takes account of the fact that the metric should be modified when
the spacetime curvature becomes comparable with $l^{-2}$. On the
other hand one can use a classical solution obtained by the
effective action of the modified gravity. This means that the Planck
length-scale, where quantum gravity effects become significant, is
much smaller than the critical scale parameter $l$. Thus, it is
reasonable to assume that the critical energy scale $\Lambda$ to be
as small as a TeV in order to solve the hierarchy problem [68-72].
This is supported by the fact that most of the phenomenological
studies of a viable fundamental theory have presumed that the
absolute maximal value of the curvature is restricted by some
fundamental value such that its corresponding characteristic energy
cannot lie far above the TeV scale [66, 73-76]. Hence, we assume
$l\sim 1$ TeV$^{-1}$. However, in a general case, $l$ is a parameter
of the corresponding UV complete theory including parameters such as
mass (and/or charge), which specify a concrete solution [67].

For $g=0$, the line element (\ref{mat:6.8}) reproduces the
Tangherlini solution of the Einstein equations. For $d=4$, one has
$g_d=g=(2Ml^2)^{\frac{1}{3}}$ and therefore the metric
(\ref{mat:6.8}) reduces to (\ref{mat:1}). The limits of large and
small $r$ of the metric function $N$ are, respectively,
\begin{equation}
\label{mat:6.10}N\approx1-\left(\frac{r_g}{r}\right)^{d-3},
\end{equation}
and
\begin{equation}
\label{mat:6.11}N\approx1-\left(\frac{r_g^{d-3}}{g_d^3}\right)r^2.
\end{equation}
These satisfy the limiting metric conditions. The critical value of
the gravitational radius $r_g^*$ can be determined by conditions
$N(r^*) = N'(r^*) = 0$ as follows:
\begin{equation}
\label{mat:6.12}
r_g^*=\left(\frac{d-1}{d-3}\right)^{\frac{1}{2}}\left(\frac{d-1}{2}\right)^{\frac{1}{d-3}}l,
\end{equation}
where the prime abbreviates $\frac{d}{dr}$. For $r_g > r_g^*$ the
line element (\ref{mat:6.8}) has two horizons, while for $r_g <
r_g^*$ there is no event horizon.

The Hawking temperature of the H-BH in $d$ dimensions takes the form
\begin{equation}
\label{mat:6.13}T_H=\frac{r_g^{d-3}}{4\pi}\frac{(d-3)r^d-2g_d^3r}{(r^{d-1}+g_d^3)^2}.
\end{equation}
The temperature vanishes in the limit $r\rightarrow r_0$ such that
for $r<r_0$, the temperature has no physical meaning. The remnant
radius $r_0$ is found to be
\begin{equation}
\label{mat:6.14}r_0=\left(\frac{2g_d^3}{d-3}\right)^{\frac{1}{d-1}}.
\end{equation}
This minimum horizon radius implies a remnant mass as follows:
\begin{equation}
\label{mat:6.15}M_0=\frac{1}{2}\left(\frac{d-1}{4^{\frac{1}{d-1}}}\right)^{\frac{1}{d-3}}
\left(\frac{g_d^3}{d-3}\right)^{\frac{1}{d-1}}.
\end{equation}

\begin{table}
\caption{\scriptsize {The remnant mass of the higher dimensional
H-BH ($\sim$ TeV) for different number of spacetime dimensions $d$
and several values of $g_d$. As the parameter $g_d$ increases the
mass of the BH remnant increases but by increasing the spacetime
dimensions the remnant mass decreases.}}
\begin{center}
\begin{tabular}{|c|c|c|c|}
\hline
\multicolumn{4}{|c|}{Higher Dimensional H-BH Remnant} \\
\hline $d$ &   $M_0$~;~$g_d=1$ & $M_0$~;~$g_d=5$  & $M_0$~;~$g_d=10$ \\
\hline$4$ & $0.94$ & $4.72$ & $9.45$\\
\hline$5$ & $0.71$ & $2.36$ & $3.98$\\
\hline$6$ & $0.62$ & $1.64$& $2.49$\\
\hline$7$ & $0.59$ & $1.31$& $1.85$\\
\hline$8$ & $0.56$ & $1.12$& $1.51$\\
\hline$9$ & $0.55$ & $1.00$& $1.30$\\
\hline$10$ & $0.54$ & $0.92$& $1.16$\\
\hline$11$ & $0.53$ & $0.86$& $1.06$\\
\hline
\end{tabular}
\end{center}
\label{tab:1.1}
\end{table}

Here, we assume that the critical energy scale $\Lambda$ is at
around the electroweak scale, i.e. $\sim$ TeV. We present the
results given in Table~\ref{tab:1.1}. We see that as $g_d$ grows the
remnant mass increases, while as the number of spacetime dimension
$d$ becomes larger the remnant mass decreases. Table~\ref{tab:1.1}
clearly shows that, for a large number of extra dimensions, the
energy scale of the minimal mass is sufficient for the energy scale
of the current runs of the LHC. As a result, if extra dimensions do
exist and if the number of spacetime dimensions becomes sufficiently
large with a sufficiently small $\Lambda$ ($\Lambda\sim 1$ TeV),
then the possible formation and detection of BHs in TeV-scale
collisions at the LHC will be enhanced.

\section{\label{sec:5}Noncommutative Hayward solution}
Our strategy here is that, firstly, the noncommutativity influences
on the spacetime of non-rotating Hayward are investigated and the
thermodynamics features of the NH-BH are determined. Afterwards, in
the later section, taking into account the Newman-Janis algorithm,
we apply the inspired noncommutativity and recompute the Hawking
temperature of the NRH-BH.

\begin{figure}[htp]
\begin{center}
\includegraphics{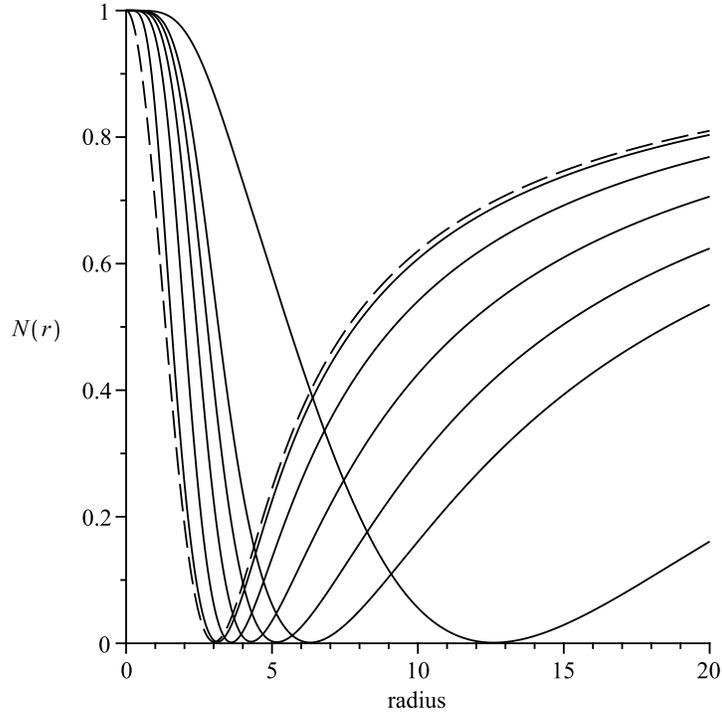}
\end{center}
\vspace{8.5 cm} \caption{\scriptsize {The temporal component of the
metric, versus the radius for different values of $g$. The figure
shows the possibility of having extremal configuration with one
degenerate event horizon at $M=M_0$ (extremal NH-BH). This shows the
existence of a minimal non-zero mass that the BH can shrink to. On
the right-hand side of the figure, from top to bottom, the solid
lines correspond to the NH-BH for $g=1.00,~ 2.00,~ 3.00,~ 4.00,~
5.00,$ and $g=10.00$, respectively. The dashed line refers to the
Schwarzschild case so that it corresponds to $g=0$.}} \label{fig:4}
\end{figure}

In accord with [77], the Newman-Janis algorithm works only for
vacuum solutions but the authors in [78] have presented a new
prescription that comprises the case of non-vanishing stress
tensors. As an introduction of the idea, let us begin by the
Schwarzschild-like form of spacetimes which explain the line
elements in the so-called Kerr-Schild classification and in the
presence of matter
\begin{equation}
\label{mat:7}ds^2=ds_M^2-\frac{h(r)}{r^2}\left(n_\alpha
dx^\alpha\right)^2,
\end{equation}
where the expression $ds_M^2$ is the Minkowski metric in a spherical
basis and $n_\alpha$ is a null vector in the coordinates of
Minkowski. The function $h(r)$ can be written as
\begin{equation}
\label{mat:8}h(r)=2m(r)r.
\end{equation}
In accordance with the Kerr-Schild decomposition, Eq.~(\ref{mat:8})
has a generic validity, thus its general form is unchanged and it is
not sensitive to various structures of the mass term. The expression
$h(r)$ for the H-BH metric is given by
\begin{equation}
\label{mat:9}h(r)=\frac{2Mr^4}{r^3+g^3}.
\end{equation}
Here, we apply the inspired noncommutative methodology [35-39] (see
also [79]). According to this method, the point-like structure of
mass, instead of being entirely localized at a point, is
characterized by a smeared structure throughout a region of linear
size $\sqrt{\theta}$. This means that the mass density of a static,
spherically symmetric, particle-like gravitational source cannot be
a delta function distribution, but will be found to be a Gaussian
distribution
\begin{equation}
\label{mat:10}\rho_{\theta}(r)=\frac{M}{(4\pi\theta)^{3/2}}e^{-\frac{r^2}{4\theta}}.
\end{equation}
The smeared mass distribution can implicity be written in terms of
the lower incomplete Gamma function,
\begin{equation}
\label{mat:11}M_\theta=\int_0^r\rho_{\theta}(r)4\pi
r^2dr=\frac{2M}{\sqrt{\pi}}\gamma\bigg(\frac{3}{2};\frac{r^2}{4\theta}\bigg).
\end{equation}
The resulting metric describing the NH-BH is given by
Eq.~(\ref{mat:1}), with the following $m(r)$ in terms of the smeared
mass distribution $M_\theta$
\begin{equation}
\label{mat:12}m(r)=M_\theta\left(\frac{r^3}{r^3+g^3}\right).
\end{equation}
The thermodynamics description of the NH-BH can now be simply
analysed by displaying the temporal component of the metric versus
the radius for an extremal BH with different values of $g$ which has
been presented in Fig.~(\ref{fig:4}) \footnote{For simplicity of
numerical calculations, we assume $\theta = 1$.}.

It is clear that the metric of the NH-BH has a coordinate
singularity at the event horizon as
\begin{equation}
\label{mat:13}r_H = 2m(r_H),
\end{equation}
with
\begin{equation}
\label{mat:14}m(r_H)=\frac{2M}{\sqrt{\pi}}\left(\frac{r_H^3}{r_H^3+g^3}\right)
\gamma\bigg(\frac{3}{2};\frac{r_H^2}{4\theta}\bigg).
\end{equation}
The analytical solution of Eq.~(\ref{mat:13}) for the horizon radius
in a closed form is not feasible, but one can solve it to obtain
$M$, which gives the mass of the NH-BH in terms of $r_H$. We find
\begin{equation}
\label{mat:15}M=\frac{r_H^3+g^3}{2r_H^2\left[{\cal{E}}\left(\frac{r_H}{2\sqrt{\theta}}\right)
-\frac{r_H}{\sqrt{\pi\theta}} e^{-\frac{r_H^2}{4\theta}}\right]},
\end{equation}
where ${\cal{E}}(n)$ is the Gaussian error function defined as $
{\cal{E}}(n)\equiv \frac{2}{\sqrt{\pi}}\int_{0}^{n}e^{-x^2}dx$. In
the limit $\theta\rightarrow0$, the Gaussian error function is equal
to one and the exponential term is reduced to zero, thus we recover
Eq.~(\ref{mat:4}). In other words, if $\sqrt{\theta}$ is too small,
the background geometry is interpreted as a smooth differential
manifold and the smeared-like mass descends to the point-like mass.
However, in the regime that noncommutative fluctuations are
important, $r\rightarrow\sqrt{\theta}$, the microstructure of
spacetime deviates considerably from the macroscopic one and
provides new physics at very short distances.

\begin{figure}[htp]
\begin{center}
\includegraphics{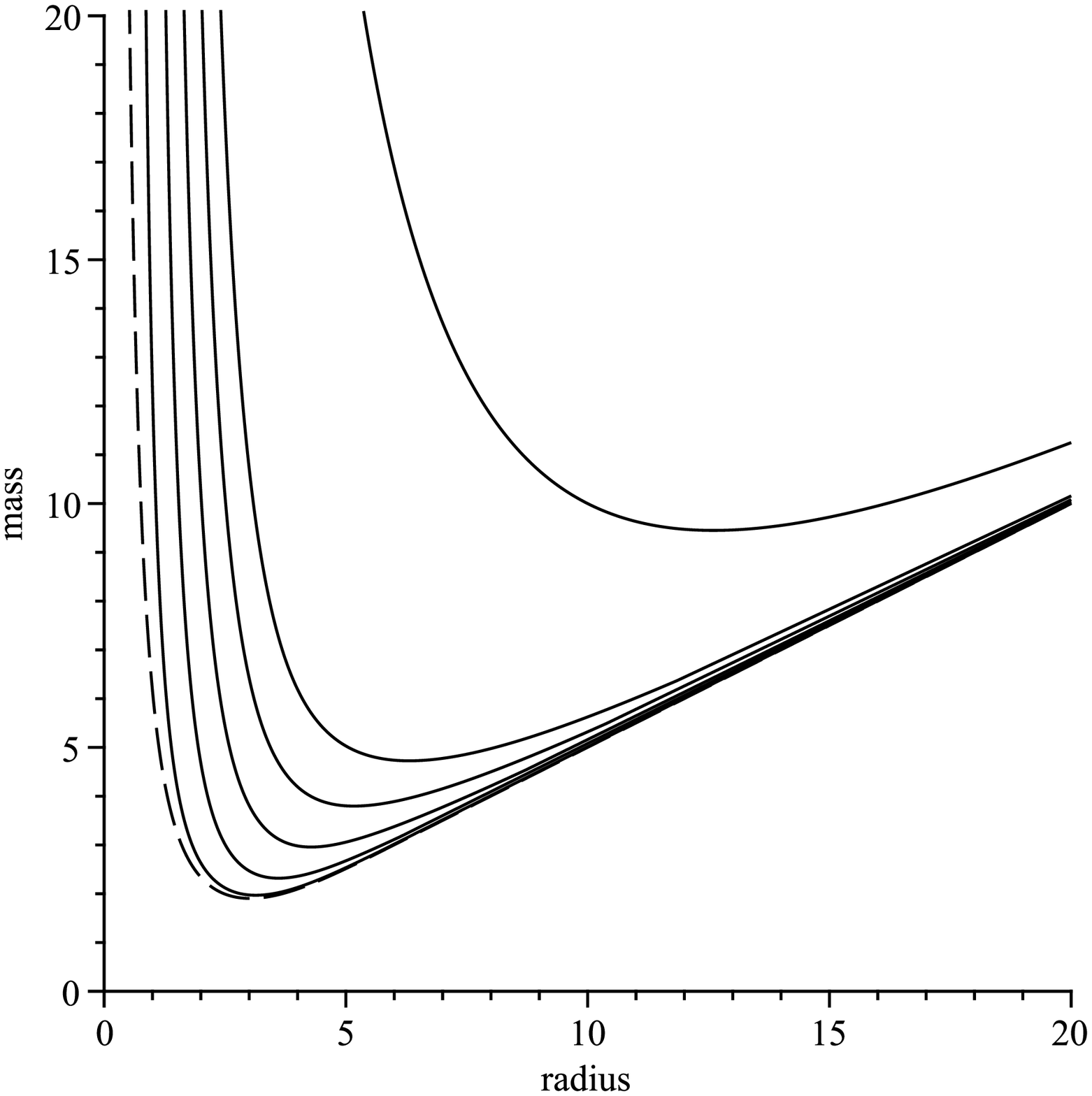}
\end{center}
\vspace{9 cm} \caption{\scriptsize {The mass of the NH-BH versus the
horizon radius for different values of $g$. On the left-hand side of
the figure, from left to right, the solid lines correspond to the
NH-BH for $g=1.00,~ 2.00,~ 3.00,~ 4.00,~ 5.00,$ and $g=10.00$,
respectively. The dashed line refers to the Schwarzschild case so
that it corresponds to $g=0$.}} \label{fig:5}
\end{figure}

The results of the numerical solution of the mass as a function of
the horizon radius are displayed in Fig.~\ref{fig:5}. According to
the numerical results it is concluded that the noncommutative
version of the mass equation (\ref{mat:15}), leads to a bigger
minimal nonzero mass at small radii in comparison with the standard
commutative version.

\begin{figure}[htp]
\begin{center}
\includegraphics{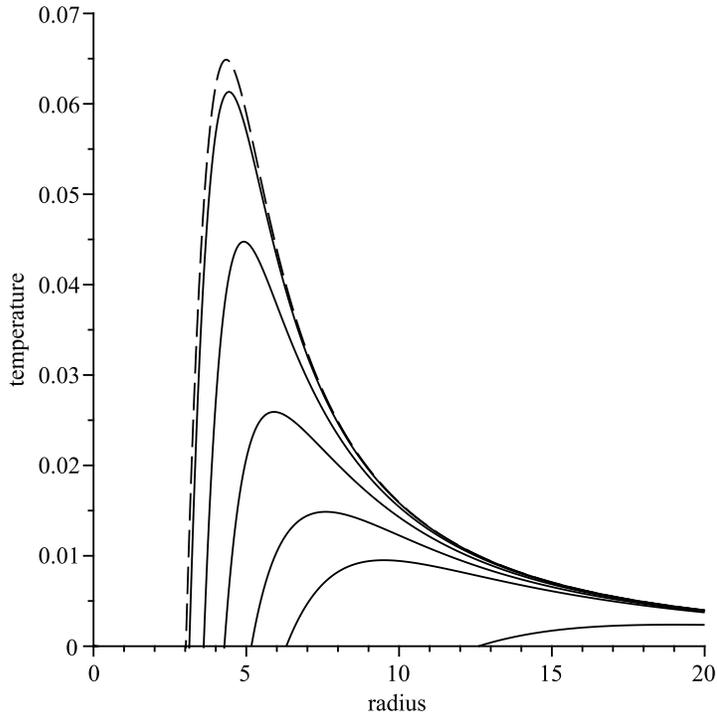}
\end{center}
\vspace{8.5 cm} \caption{\scriptsize {The Hawking temperature versus
the horizon radius. We have set $M=10.00$. On the left-hand side of
the figure, from left to right, the solid lines correspond to the
NH-BH for $g=1.00,~ 2.00,~ 3.00,~ 4.00,~ 5.00,$ and $g=10.00$,
respectively. The dashed line refers to the Schwarzschild case so
that it corresponds to $g=0$.}} \label{fig:6}
\end{figure}

The Hawking temperature of the NH-BH can be written as
\begin{eqnarray}
\label{mat:16}T_H=\frac{M}{4\sqrt{(\pi\theta)^3}\left(r_H^3+g^3\right)^2}
\bigg[4r_H\sqrt{\pi\theta^3}
\left(\frac{r_H^3}{2}-g^3\right){\cal{E}}\left(\frac{r_H}{2\sqrt{\theta}}\right)
-r_H^2e^{-\frac{r_H^2}{4\theta}}\nonumber
\\ \times\left(r_H^5+2r_H^3\theta+r_H^2g^3-4\theta
g^3\right)\bigg].
\end{eqnarray}
In the commutative version and for $g=0$, the Gauss error function
is unity and the exponential term is zero, so we retrieve the
Hawking temperature of a Schwarzschild BH. The numerical result of
the NH-BH temperature in terms of the horizon radius is shown in
Fig.~\ref{fig:6}. From the figure we see that the maximum
temperature decreases with raising the parameter $g$. As a result,
the size and the mass of the NH-BH remnant at the ultimate phase of
the evaporation is bigger in comparison with the noncommutative
Schwarzschild one.

For more specifications, we present Table~\ref{tab:2} that is
similar to Table~\ref{tab:1}. From Table~\ref{tab:2} we see that as
$g$ grows both the remnant mass and the remnant radius are increased
which finally, in the limit $g\gg1$, yields a proportional
relationship $g\propto M_0\propto r_0$. A comparison between the
final stages of the evaporation for the noncommutative Schwarzschild
BH and the NH-BH shows that raising the size and the mass of the
remnant and also getting a colder BH is affected by an increase in
the parameter $g$. In addition, as can be seen from
Figs.~(\ref{fig:4}),~(\ref{fig:5}), ~(\ref{fig:6}) and
Table~\ref{tab:2} the noncommutative coordinates yields a bigger
remnant and also a colder BH at small radii compared to its
commutative case.

\begin{table}
\caption{\scriptsize {The table in the upper place shows the remnant
mass, the remnant radius and the maximum temperature of the
noncommutative Schwarzschild BH, while the table below shows them
for the NH-BH with different values of $g$.}}
\begin{center}
\begin{tabular}{|c|c|c|}
\hline
\multicolumn{3}{|c|}{Noncommutative Schwarzschild BH} \\
\hline  Remnant Mass & Remnant Radius & Maximum Temperature \\
\hline $M_0\approx1.90$ & $r_0\approx3.02$ & $T_H(max)\approx0.065$\\
\hline
\end{tabular}
{\vspace{0.1cm}}
\begin{tabular}{|c|c|c|c|}
\hline
\multicolumn{4}{|c|}{NH-BH} \\
\hline Free Parameter & Remnant Mass & Remnant Radius & Maximum Temperature \\
\hline$g=1.00$ & $M_0\approx1.96$ & $r_0\approx3.13$ & $T_H(max)\approx0.062$\\
\hline$g=2.00$ & $M_0\approx2.31$ & $r_0\approx3.60$ & $T_H(max)\approx0.045$\\
\hline$g=3.00$ & $M_0\approx2.95$ & $r_0\approx4.28$ & $T_H(max)\approx0.026$\\
\hline$g=4.00$ & $M_0\approx3.79$ & $r_0\approx5.16$ & $T_H(max)\approx0.015$\\
\hline$g=5.00$ & $M_0\approx4.72$ & $r_0\approx6.31$ & $T_H(max)\approx0.009$\\
\hline$g=10.00$ & $M_0\approx9.45$ & $r_0\approx12.60$ & $T_H(max)\approx0.002$\\
\hline
\end{tabular}
\end{center}
\label{tab:2}
\end{table}

According to our results, for $g<5$ the minimum required energy for
the formation of NH-BHs at particle colliders such as LHC will be
larger compared to the H-BH case. This is indeed a consequence of
noncommutative effects and may be interpreted as an indication of a
suppression of the BH production arisen from the local fluctuations
of the geometry at short distances. This is in agreement with the
results obtained in the context of gravity's rainbow [66]. The
authors in Ref.~[66] have proposed this as a possible explanation
for the absence of BHs at the LHC. In addition, they have found that
a remnant depends critically on the structure of the rainbow
functions [80]. They have argued that, using the framework of
gravity's rainbow, a remnant is formed for all black objects in such
a way that it is a model-independent phenomenon.

Given that the physical behavior of the H-BH is qualitatively the
same with or without noncommutativity, one might ask the question
"is there any reason to introduce a smearing of the source and/or
the Hayward's parameter?". In order to answer this question we
should explain that the temperature is considerably different and is
lower in the case of the noncommutativity. By changing
noncommutative's parameter and keeping the Hayward's parameter to be
constant, the results are qualitatively similar. Nevertheless, in
principle, there are two good reasons to show that there are
fundamental differences between two cases. First, the spacetime
noncommutativity does not depend on the curvature, but is an
intrinsic property of the manifold itself even in the absence of
gravity which is denoted by the parameter $\theta$ and can eliminate
some kind of divergences which appear in general relativity. Hence,
if any effect is produced by the noncommutativity it must appear
also in weak fields. Second, the concept of "weak" or "strong" field
is sensible only if one compares the field strength with a proper
scale. In the gravitation theory, we have a natural and unique
scale, that is the Planck scale. Therefore, the gravitational field
strength can still be considered "weak" even near a BH, with respect
to the Planck scale. This issue justifies the adoption of linearized
field equations as a temporary laboratory to test the effect of
noncommutativity until the horizon radius is larger than the Planck
length [81].

From the other point of view, the H-BH is a regular solution of a
modified Einstein equation, and it is also found in the Einstein
gravity coupled with NED. It is well-known that in the near horizon
of a BH, the quantum effect becomes important because of the strong
gravity, therefore the geometry of the spacetime can be modified
from the quantum effect at the near horizon and the intrinsic
singularity inside the BH can be eliminated. One could expect that
the metric of the BH is modified in the near horizon region due to
the quantum effect. The extent of the deviation from the standard
solution of Einstein equations is denoted by the free parameter $g$.
Hence, the parameter $g$ can describe how much the quantum effect
near the horizon affects the deviation from the standard energy
level and the radiation. Along this line of reasoning we take them
as two distinct situations.

\begin{figure}[htp]
\begin{center}
\includegraphics{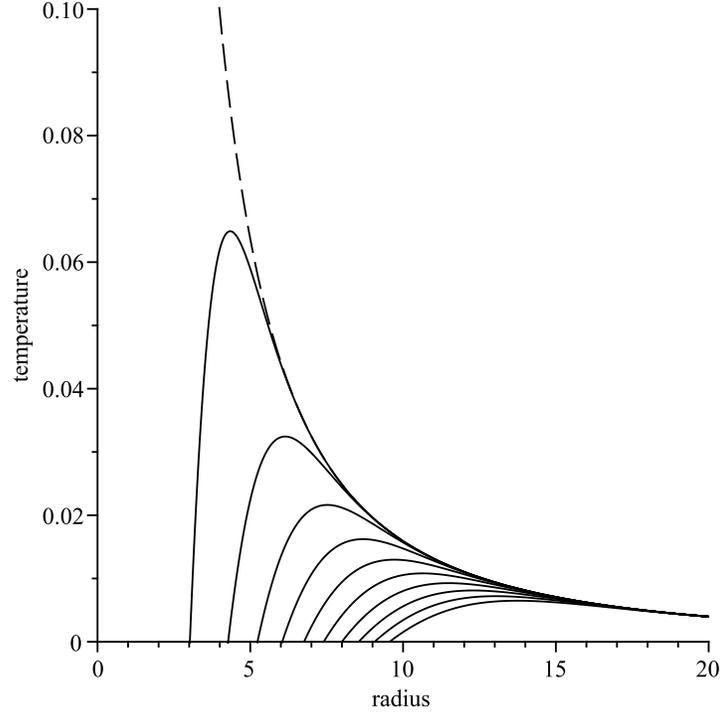}
\end{center}
\vspace{8.5 cm} \caption{\scriptsize {The Hawking temperature versus
the horizon radius. We have set $M=10.00$. On the left-hand side of
the figure, from left to right, the solid lines correspond to the
noncommutative Schwarzschild BH for $\theta=1.00,~ 2.00,~ 3.00,~
\cdots ~ ,~10.00$, respectively. The dashed line refers to the
commutative Schwarzschild BH so that it corresponds to $\theta=0$.}}
\label{fig:7}
\end{figure}

\begin{figure}[htp]
\begin{center}
\includegraphics{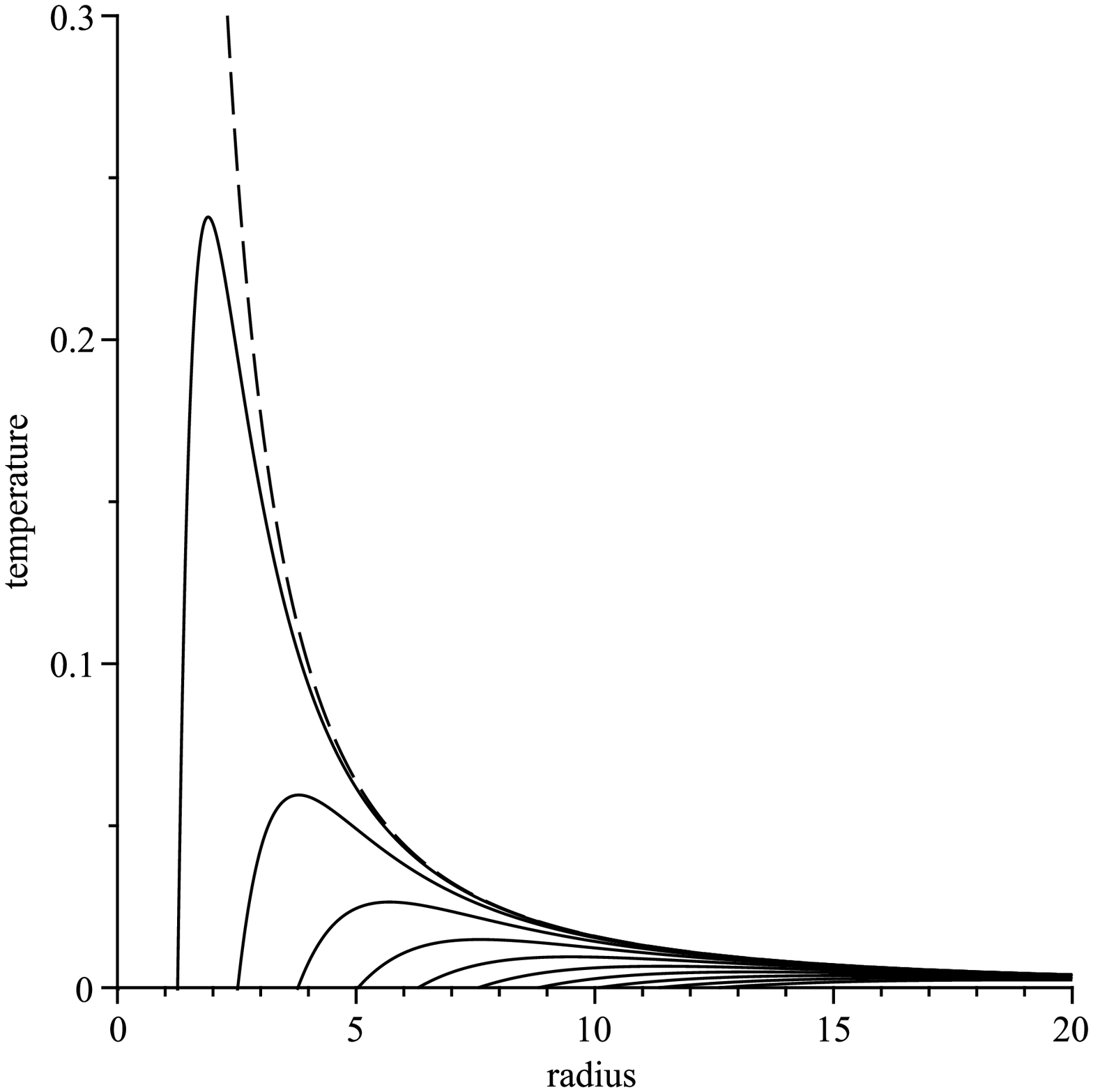}
\end{center}
\vspace{8.5 cm} \caption{\scriptsize {The Hawking temperature versus
the horizon radius. We have set $M=10.00$. On the left-hand side of
the figure, from left to right, the solid lines correspond to the
H-BH for $g=1.00,~ 2.00,~ 3.00,~ \cdots ~,~10.00$, respectively. The
dashed line refers to the Schwarzschild BH so that it corresponds to
$g=0$.}} \label{fig:8}
\end{figure}

For further specifications, the numerical result of the temperature
in terms of the horizon radius for the noncommutative Schwarzschild
and Hayward BHs are shown in Figs.~\ref{fig:7} and \ref{fig:8},
respectively. We just change the value of $\theta$ for $g=0$. From
the figures we see that the maximum temperature decreases
considerably with increasing the parameter $\theta$ compared to the
results obtained just by tuning $g$. As a result, the size and the
mass of the NH-BH remnant at the ultimate phase of the evaporation
is bigger for small amounts of $\theta$, while is smaller for large
amounts of $\theta$ in comparison with the case that we change the
Hayward's parameter and keeping $\theta$ equal to zero. In fact,
noncommutative's parameter is more sensitive to small radii, while
Hayward's parameter has a linear relationship with the remnant
radius.

It may be noted that as another striking example of regular BHs, if
the Bardeen solution is chosen, solely the mass term will be
changed, however the general properties will be directed to entirely
comparable consequences to those above [82].

\section{\label{sec:6}Noncommutative rotating Hayward solution}
For spinning solution, we apply the Newman-Janis algorithm, and
assuming that the mass term $m(r)$ is not affected by the
complexification $r\rightarrow r'=r+ia\cos\vartheta$, the general
form of the Kerr-Schild decomposition (\ref{mat:7}) holds
\begin{equation}
\label{mat:17}ds^2=ds_M^2-\frac{h(r)}{r'\bar{r}'}\left(n_\alpha
dx^\alpha\right)^2,
\end{equation}
where $n_\alpha$ is written in spheroidal coordinates and $h(r)$ is
unaltered by expressing $m(r')$ as $m(\textrm{Re}(r'))=m(r)$. In
fact, even with changing the symmetry from a spherically symmetric
geometry to an axially symmetric geometry, the formal structure of
the Kerr-Schild solution does not change. This is consistent with
the solution of type-I or the first-class solution of the RH-BH in
[55], i.e. the complexification of the $\frac{1}{r}$ term as in
Schwarzschild, without changing the mass term.

Now, with the above explanation, one can obtain the line element of
RNH-BH in Boyer-Lindquist coordinates
\begin{eqnarray}
\label{mat:18}ds^2=
\frac{\Delta-a^2\sin^2\vartheta}{\Sigma}dt^2-\frac{\Sigma}{\Delta}dr^2-\Sigma
d\vartheta^2+2a\sin^2\vartheta\left(1-\frac{\Delta-a^2\sin^2\vartheta}{\Sigma}\right)dtd\phi\nonumber
\\
-\sin^2\vartheta\left[\Sigma+a^2\sin^2\vartheta\left(2-\frac{\Delta-a^2\sin^2\vartheta}{\Sigma}\right)\right]d\phi^2,
\end{eqnarray}
where $\Delta:=r^2-2m(r)r+a^2$ (with $m(r)$ given by
Eq.~(\ref{mat:12})) and $\Sigma:=r^2+a^2\cos^2\vartheta$.

\begin{figure}[htp]
\begin{center}
\includegraphics{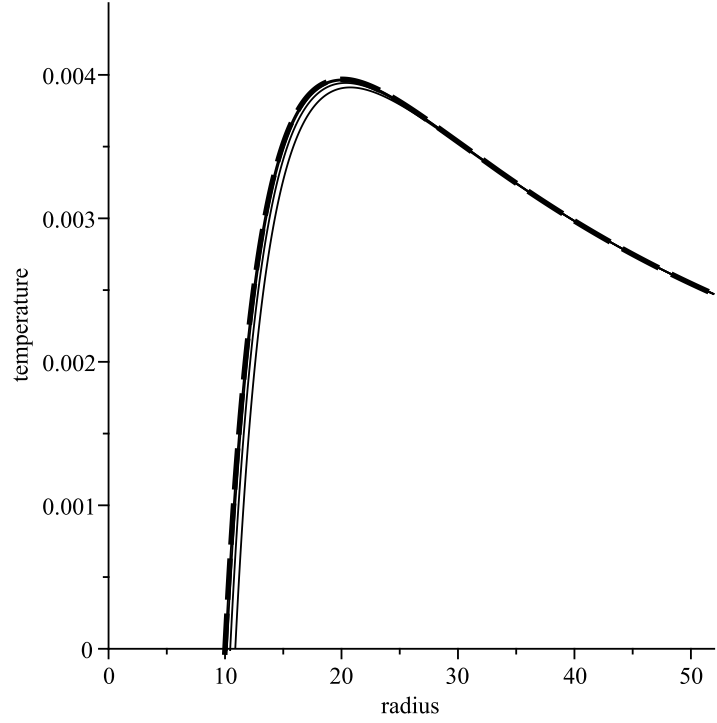}
\end{center}
\vspace{8.5 cm} \caption{\scriptsize {The temperature $T_H$ versus
the outer horizon radius, $r_+$. We have set $M=10.00$ and $a=1.00$.
On the left-hand side of the figure, from left to right, the solid
lines correspond to the NRH-BH for $g=1.00,~ 2.00,~ 3.00,~ 4.00,~
5.00,$ and $g=10.00$, respectively. The dashed line refers to the
noncommutative Kerr BH so that it corresponds to $g=0$. It is clear
that the curves are not so sensitive to $g$.}} \label{fig:9}
\end{figure}

The Hawking temperature of the NRH-BH is then found to be
$$T_H=\frac{1}{4\pi(r_+^2+a^2)}\frac{d\Delta}{dr}\Bigg|_{r=r_+}$$$$=-\frac{1}{4\sqrt{(\pi\theta)^3}(r_+^3+g^3)^2(r_+^2+a^2)}
\bigg[8\sqrt{\pi\theta^3}
Mr_+^3\left(\frac{r_+^3}{4}+g^3\right){\cal{E}}\left(\frac{r_+}{2\sqrt{\theta}}\right)$$
\begin{equation}
\label{mat:19} +\left(Mr_+^9-2M\theta r_+^7+Mg^3r_+^6-8M\theta
g^3r_+^4\right)e^{-\frac{r_+^2}{4\theta}}-2\sqrt{\pi\theta^3}r_+(r_++g)^2(r_+^2-gr_++g^2)^2\Bigg].
\end{equation}
Note that, for the commutative case and for $g=a=0$, the function
${\cal{E}}\left(\frac{r_+}{2\sqrt{\theta}}\right)$ becomes one and
the exponential term is zero, but the last term in
Eq.~(\ref{mat:19}) which is independent of the mass $M$ will be
reduced to $\frac{1}{2\pi r_+}$. Hence, one retrieves the standard
result
\begin{equation}
\label{mat:20}T_H=-\frac{M}{2\pi r_+^2}+\frac{1}{2\pi
r_+}=\frac{1}{4\pi r_+},
\end{equation}
where in this case $r_+=r_H=2M$. Finally, the numerical result of
the Hawking temperature as a function of the outer horizon radius
(Eq.~(\ref{mat:19})) is shown in Fig.~\ref{fig:9}. In accord with
the figure, the size and the mass of the NRH-BH remnant at the final
stage of the evaporation increase with increasing the parameter $g$.
However, in the rotating case it is not so sensitive to $g$ with
respect to the non-rotating one.

\begin{figure}[htp]
\begin{center}
\includegraphics{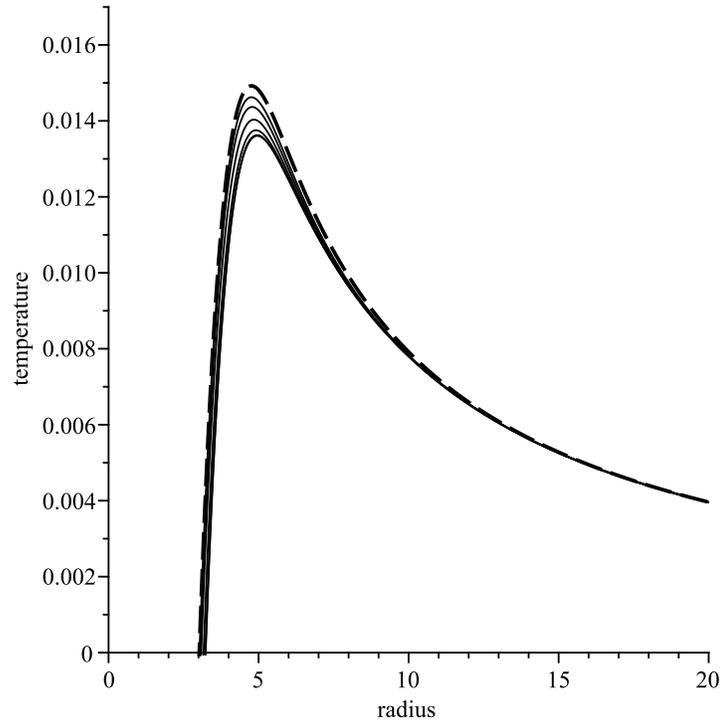}
\end{center}
\vspace{8.5 cm} \caption{\scriptsize {The temperature $T_H$ versus
the outer horizon radius, $r_+$. We have set $M=10.00$ and $a=1.00$.
The solid lines through the center of the figure, from top to
bottom, correspond to the RH-BH for $g=1.00,~ 2.00,~ 3.00,~ 4.00,~
5.00,$ and $g=10.00$, respectively. The dashed line refers to the
Kerr BH so that it corresponds to $g=0$.}} \label{fig:10}
\end{figure}

In order to compare the noncommutative results with the commutative
case ($\theta\rightarrow0$), we display the plot for the temperature
of the RH-BH as a function of $r_+$ (see Fig.~\ref{fig:10}). One can
see that the noncommutative effects are caused to have a larger size
and mass of the remnant in addition to a colder BH. Furthermore, it
is reasonable to expect that the enlarging the amounts of the
remnant mass is the role of the rotation as well.

Generally, in Figs.~\ref{fig:9} and ~\ref{fig:10}, we see that the
feature of the temperature is nearly similar to that of the
nonrotating case. After a temperature peak, the NRH-BH calms down to
a zero temperature as a NRH-BH remnant at the final phase of its
evaporation. The size and the mass of this remnant become larger
with respect to the nonrotating solution and this is due to the fact
that the rotational kinetic energy is kept in the final format. As a
consequence, the noncommutative effects and the rotation factor can
raise the minimum value of energy for the possible production of BHs
in TeV-scale collisions at particle colliders, so that the
possibility for the formation and detection of BHs would be reduced.

\section{\label{sec:7}Summary}
We have proposed that the final phase of the BH evaporation is a
stable remnant. In this study, the H-BH as a most popular model of
regular BHs has been chosen. We have verified that the exact H-BH
solution extracted from the theory of NED coupled with gravity
satisfies sufficient conditions for the linear stability with
respect to arbitrary linear fluctuations. The thermodynamics
features of its non-rotating and rotating solution in the presence
of an inspired model of noncommutative geometry have been analysed.
The effect of this inspired noncommutativity in the microscopic
feature of spacetime is that all point structures are replaced by
structures smeared via such an inspired microstructure. We have
explored the effect that the deformation of a point mass has on the
thermodynamics of H-BH solutions. Thus, the corrections to the
Hawking temperature via such a modification of the theory have been
found. Finally we have extended our analysis to the thermodynamic
properties of noncommutative spinning solutions, providing their
Hawking temperature. It is concluded that, the noncommutative
effects cause an increasing size and mass of the remnant and also
making the BH to be colder in the small radii regime as the free
parameter $g$ increases. As a result, the noncommutative effects can
enhance the minimum required energy for the creation of such BHs in
the present experimental attempts at the LHC. This may reduce the
possibility for the formation and detection of BHs in TeV-scale
collisions at particle colliders. However, if we have enough chance
to have large extra dimensions, then it may make the BH production
experimentally accessible at colliders.

\section*{Acknowledgments}
Financial support by Lahijan Branch, Islamic Azad University Grant
No. 17.20.5.3517 is gratefully acknowledged. The authors thank to C.
Bambi for valuable suggestions. Also, the authors would like to thank
the referee for useful comments.\\


\begin{thebibliography}{99}
\bibitem{1}
See for instance, S. W. Hawking and G. Ellis, {\it The Large Scale
Structure of Space-Time}, Cambridge University Press, Cambridge,
1973
\bibitem{2}
See for a review, S. Ansoldi, [arXiv:0802.0330]
\bibitem{3}
J. M. Bardeen, in Conference Proceedings of GR5 (Tbilisi, USSR,
1968), p. 174
\bibitem{4}
S. A. Hayward, Phys. Rev. Lett. {\bf96}, 031103 (2006),
[arXiv:gr-qc/0506126]
\bibitem{5}
J. C. S. Neves and A. Saa, Phys. Lett. B {\bf734}, 44 (2014),
[arXiv:1402.2694]
\bibitem{6}
K. Lin, J. Li and S. Yang, Int. J. Theor. Phys. {\bf52}, 3771 (2013)
\bibitem{7}
G. Abbas and U. Sabiullah, Astrophys. Space Sci. {\bf352}, 769
(2014), [arXiv:1406.0840]
\bibitem{8}
U. Debnath, Eur. Phys. J. C {\bf75}, 129 (2015), [arXiv:1503.01645]
\bibitem{9}
B. Pourhassan, M. Faizal and Ujjal Debnath,  Eur. Phys. J. C
{\bf76}, 145 (2016), [arXiv:1603.01457]
\bibitem{10}
M. Azreg-A\"{\i}nou, Phys. Rev. D {\bf90}, 064041 (2014),
[arXiv:1405.2569]
\bibitem{11}
V. P. Frolov, JHEP {\bf05}, 49 (2014), [arXiv:1402.5446]
\bibitem{12}
E. O. Kahya, M. Khurshudyan, B. Pourhassan, R. Myrzakulov and A.
Pasqua, Eur. Phys. J. C {\bf75}, 43 (2015), [arXiv:1402.2592]
\bibitem{13}
T. D. Lorenzo, C. Pacilioy, C. Rovelli and S. Speziale, Gen. Rel.
Grav. {\bf47}, 41 (2015), [arXiv:1412.6015]
\bibitem{14}
T. D. Lorenzo, A. Giusti and S. Speziale, Gen. Rel. Grav. {\bf48},
31 (2016), [arXiv:1510.08828]
\bibitem{15}
E. Ay\'{o}n-Beato and A. Garc\'{\i}a, Phys. Rev. Lett. {\bf80}, 5056
(1998), [arXiv:gr-qc/9911046]
\bibitem{16}
E. Ay\'{o}n-Beato and A. Garc\'{\i}a, Phys. Lett. B {\bf464}, 25
(1999), [arXiv:hep-th/9911174]
\bibitem{17}
E. Ay\'{o}n-Beato and A. Garc\'{\i}a, Gen. Relativ. Gravit. {\bf31},
629 (1999), [arXiv:gr-qc/9911084]
\bibitem{18}
E. Ay\'{o}n-Beato and A. Garc\'{\i}a, Phys. Lett. B {\bf493}, 149
(2000), [arXiv:gr-qc/0009077]
\bibitem{19}
M. Born and L. Infeld, Proc. R. Soc. London A {\bf144}, 425 (1934)
\bibitem{20}
E. S. Fradkin and A. A. Tseytlin, Phys. Lett. B {\bf163}, 123 (1985)
\bibitem{21}
A. A. Tseytlin, Nucl. Phys. B {\bf276}, 391 (1986)
\bibitem{22}
N. Seiberg and E. Witten, JHEP {\bf9909}, 032 (1999),
[arXiv:hep-th/9908142]
\bibitem{23}
E. L. B. Junior, M. E. Rodrigues, and M. J. S. Houndjo, J. Cosmol.
Astropart. Phys. {\bf10}, 060 (2015), [arXiv:1503.07857]
\bibitem{24}
Z-Y. Fan, Eur. Phys. J. C {\bf77}, 266 (2017), [arXiv:1609.04489]
\bibitem{25}
Z-Y. Fan and X. Wang, Phys. Rev. D {\bf94}, 124027 (2016),
[arXiv:1610.02636]
\bibitem{26}
B. Toshmatov, Z. Stuchl\'{\i}k and B. Ahmedov, Phys. Rev. D {\bf95},
084037 (2017), [arXiv:1704.07300]
\bibitem{27}
E. Ay\'{o}n-Beato and A. Garc\'{\i}a, Gen. Rel. Grav. {\bf37}, 635
(2005), [arXiv:hep-th/0403229]
\bibitem{28}
K. A. Bronnikov and J. C. Fabris, Phys. Rev. Lett. {\bf96}, 251101
(2006), [arXiv:gr-qc/0511109]
\bibitem{29}
W. Berej, J. Matyjasek, D. Tryniecki and M. Woronowicz, Gen. Rel.
Grav. {\bf38}, 885 (2006), [arXiv:hep-th/0606185]
\bibitem{30}
M. Azreg-A\"{\i}nou, Phys. Lett. B {\bf730}, 95 (2014),
[arXiv:1401.0787]
\bibitem{31}
L. Balart and E. C. Vagenas, Phys. Lett. B {\bf730}, 14 (2014),
[arXiv:1401.2136]
\bibitem{32}
I. Radinschi, F. Rahaman, T. Grammenos, A. Spanou and S. Islam, Adv.
Math. Phys. {\bf2015}, 530281 (2015), [arXiv:1404.6410]
\bibitem{33}
I. Dymnikova and E. Galaktionov, Class. Quant. Grav. {\bf32}, 165015
(2015), [arXiv:1510.01353]
\bibitem{34}
E. Chaverra, J. C. Degollado, C. Moreno and O. Sarbach, Phys. Rev. D
{\bf93}, 123013 (2016), [arXiv:1605.04003]
\bibitem{35}
P. Nicolini, A. Smailagic and E. Spallucci, Phys. Lett. B {\bf632},
547 (2006), [arXiv:gr-qc/0510112]
\bibitem{36}
S. Ansoldi, P. Nicolini, A. Smailagic and E. Spallucci, Phys. Lett.
B {\bf645}, 261 (2007), [arXiv:gr-qc/0612035]
\bibitem{37}
K. Nozari and S. H. Mehdipour, Class. Quant. Grav. {\bf25}, 175015
(2008), [arXiv:0801.4074]
\bibitem{38}
See for a review, P. Nicolini, Int. J. Mod. Phys. A {\bf24}, 1229
(2009), [arXiv:0807.1939]
\bibitem{39}
S. H. Mehdipour, Phys. Rev. D {\bf81}, 124049 (2010),
[arXiv:1006.5215]
\bibitem{40}
P. Chen, Y. C. Ong and D. H. Yeom, Phys. Rep. {\bf603}, 1 (2015),
[arXiv:1412.8366]
\bibitem{41}
C. Bambi, Mod. Phys. Lett. A {\bf26}, 2453 (2011), [arXiv:1109.4256]
\bibitem{42}
C. Bambi, Phys. Lett. B {\bf730}, 59 (2014), [arXiv:1401.4640]
\bibitem{43}
U. Debnath, Eur. Phys. J. C {\bf75}, 129 (2015), [arXiv:1503.01645]
\bibitem{44}
C. Bambi, Z. Cao and L. Modesto, Phys. Rev. D {\bf95}, 064006
(2017), [arXiv:1701.00226]
\bibitem{45}
B. P. Abbott {\it et al.} (LIGO Scientific Collaboration and Virgo
Collaboration), Phys. Rev. Lett. {\bf116}, 061102 (2016),
[arXiv:1602.03837]
\bibitem{46}
B. P. Abbott {\it et al.} (LIGO Scientific Collaboration and Virgo
Collaboration), Phys. Rev. Lett. {\bf116}, 241102 (2016),
[arXiv:1602.03840]
\bibitem{47}
B. P. Abbott {\it et al.} (LIGO Scientific Collaboration and Virgo
Collaboration), Phys. Rev. Lett. {\bf116}, 131102 (2016),
[arXiv:1602.03847]
\bibitem{48}
B. P. Abbott {\it et al.} (LIGO Scientific Collaboration and Virgo
Collaboration), Phys. Rev. Lett. {\bf116}, 241103 (2016),
[arXiv:1606.04855]
\bibitem{49}
B. Gwak, [arXiv:1703.10154]
\bibitem{50}
E. F. Eiroa and C. M. Sendra, Class. Quant. Grav. {\bf28}, 085008
(2011), [arXiv:1011.2455]
\bibitem{51}
J. Schee, Z. Stuchl\'{\i}k, J. Cosmol. Astropart. Phys. {\bf06}, 048
(2015), [arXiv:1501.00835]
\bibitem{52}
A. Abdujabbarov, M. Amir, B. Ahmedov and S. G. Ghosh, Phys. Rev. D
{\bf93}, 104004 (2016), [arXiv:1604.03809]
\bibitem{53}
S-S. Zhao and Y. Xie, [arXiv:1704.02434]
\bibitem{54}
E. T. Newman and A. I. Janis, J. Math. Phys. {\bf6}, 915 (1965)
\bibitem{55}
C. Bambi and L. Modesto, Phys. Lett. B {\bf721}, 329 (2013),
[arXiv:1302.6075]
\bibitem{56}
Y. S. Myung, Y-W. Kim, and Y. J. Park, Phys. Lett. B {\bf656}, 221
(2007), [arXiv:gr-qc/0702145]
\bibitem{57}
Y. S. Myung, Y-W. Kim and Y-J. Park, Gen. Rel. Grav. {\bf41}, 1051
(2009), [arXiv:0708.3145]
\bibitem{58}
C. Moreno and O. Sarbach, Phys. Rev. D {\bf67}, 024028 (2003),
[arXiv:gr-qc/0208090]
\bibitem{59}
N. Breton, Phys. Rev. D {\bf72}, 044015 (2005),
[arXiv:hep-th/0502217]
\bibitem{60}
B. Koch, M. Bleicher, and S. Hossenfelder, JHEP {\bf0510}, 053
(2005), [arXiv:hep-ph/0507138]
\bibitem{61}
J. L. Hewett, B. Lillie, and T. G. Rizzo, Phys. Rev. Lett. {\bf95},
261603 (2005), [arXiv:hep-ph/0503178]
\bibitem{62}
G. L. Alberghi, R. Casadio, and A. Tronconi, J. Phys. G {\bf34}, 767
(2007), [arXiv:hep-ph/0611009]
\bibitem{63}
L. Modesto, Proceedings of the XVII SIGRAV Conference, Turin,
September 4-7, 2006, [arXiv:hep-th/0701239]
\bibitem{64}
G. Amelino-Camelia, J. Ellis, N. E. Mavromatos and D. V. Nanopoulos,
Int. J. Mod. Phys. A {\bf12}, 607 (1997), [arXiv:hep-th/9605211]
\bibitem{65}
A. F. Ali, Phys. Rev. D {\bf89}, 104040 (2014), [arXiv:1402.5320]
\bibitem{66}
A. F. Ali, M. Faizal and M. M. Khalil, Phys. Lett. B {\bf743}, 295
(2015), [arXiv:1410.4765]
\bibitem{67}
V. P. Frolov, Phys. Rev. D {\bf94}, 104056 (2016),
[arXiv:1609.01758]
\bibitem{68}
I. Antoniadis, Phys. Lett. B {\bf246}, 377 (1990)
\bibitem{69}
J. D. Lykken, Phys. Rev. D {\bf54}, 3693 (1996),
[arXiv:hep-th/9603133]
\bibitem{70}
E. Witten, Nucl. Phys. B {\bf471}, 135 (1996),
[arXiv:hep-th/9602070]
\bibitem{71}
N. Arkani-Hamed, S. Dimopoulos and G. R. Dvali, Phys. Lett. B
{\bf429}, 263 (1998), [arXiv:hep-ph/9803315]
\bibitem{72}
L. Randall and R. Sundrum, Phys. Rev. Lett. {\bf83}, 3370 (1999),
[arXiv:hep-ph/9905221]
\bibitem{73}
I. Hinchliffe, N. Kersting and Y. L. Ma, Int. J. Mod. Phys. A
{\bf19}, 179 (2004), [arXiv:hep-ph/0205040]
\bibitem{74}
J. M. Conroy, H. J. Kwee and V. Nazaryan, Phys. Rev. D {\bf68},
054004 (2003), [arXiv:hep-ph/0305225]
\bibitem{75}
P. Schupp, J. Trampetic, J. Wess and G. Raffelt, Eur. Phys. J. C
{\bf36}, 405 (2004), [arXiv:hep-ph/0212292]
\bibitem{76}
K. Nozari and S. H. Mehdipour, JHEP {\bf03}, 061 (2009),
[arXiv:0902.1945]
\bibitem{77}
S. P. Drake and P. Szekeres, Gen. Rel. Grav. {\bf32}, 445 (2000),
[arXiv:gr-qc/9807001]
\bibitem{78}
L. Modesto and P. Nicolini, Phys. Rev. D {\bf82}, 104035 (2010),
[arXiv:1005.5605]
\bibitem{79}
I. Arraut, D. Batic and M. Nowakowski, J. Math. Phys. {\bf51},
022503 (2010), [arXiv:1001.2226]
\bibitem{80}
A. F. Ali, M. Faizal and M. M. Khalil, Nucl. Phys. B {\bf894}, 341
(2015), [arXiv:1410.5706]
\bibitem{81}
P. Nicolini, J. Phys. A {\bf38}, L631 (2005), [arXiv:hep-th/0507266]
\bibitem{82}
S. H. Mehdipour and M. H. Ahmadi, Astrophys. Space Sci. {\bf361},
314 (2016), [arXiv:1604.06272]

\end{thebibliography}
\end{document}